\documentclass[11pt,twoside]{article}
\usepackage{./asp2014}

\aspSuppressVolSlug
\resetcounters

\begin{document}

\title{Analysis methods of the flow field around decaying sunspots}
\author{Hanna Strecker$^1$, and Nazaret Bello Gonz\'alez$^2$
\affil{$^1$Kiepenheuer-Institut für Sonnenphysik, Freiburg, Germany; \email{strecker@leibniz-kis.de}}
\affil{$^2$Kiepenheuer-Institut für Sonnenphysik, Freiburg, Germany; \email{nbello@leibniz-kis.de}}}

\paperauthor{Hanna Strecker}{strecker@leibniz-kis.de}{ORCID_Or_Blank}{Kiepenheuer-Institut für Sonnenphysik}{Author1 Department}{Freiburg}{State/Province}{79104}{Freiburg}
\paperauthor{Nazaret Bello Gonz\'alez}{nebllo@leibniz-kis.de}{ORCID_Or_Blank}{Kiepenheuer-Institut für Sonnenphysik}{Author2 Department}{Freiburg}{State/Province}{79104}{Freiburg}

\begin{abstract}
The moat flow, a radial outflow surrounding fully fledged sunspots, is a well characterised phenomenon. Nevertheless, its origin and especially its relation to the penumbra is still a controversial topic. We investigate the evolution of the horizontal velocity of the flow around sunspots over several days during sunspot decay. SDO/HMI Doppler maps, which allow for the continuous observation of an active regions, are used. We describe the analysis method used to retrieve the horizontal velocity of the flow field for different positions on the solar disc. For that purpose, several large and small scale flow patterns, like, e.g., differential rotation, the centre-to-limb variation in the convective blueshift and a residual pattern caused by instrumental effects, have to be taken into account in order to properly measure the horizontal velocity of the flow field surrounding the sunspots. We find that the flow field around sunspots with fully developed penumbra has a decreasing velocity profile with increasing distances to the sunspot, as already found by other authors. Most important, the velocity amplitude decreases and the profile changes as the penumbra dissolves and the sunspots decay. Our findings confirm the related disappearance of the moat flow with penumbra. Yet, we observe a remnant outflow after the penumbra disappears, which hints towards the possible overtaken of the moat flow by a supergranular flow in decaying sunspots.
\end{abstract}
%%%%%%%%%%%%%%%%%%%%%%
\section{Introduction}
Fully evolved sunspots with penumbra are know to be surrounded by a flow field --~the moat flow. It is characterised by a horizontal-radially outward directed flow of gas away from the spot. \citet{2013A&A...551A.105L} found for 31 H-class sunspots extension values of the flow field from the penumbral boundary in the range of 5-15~Mm with a mean value of 9.2~Mm. The moat flow shows horizontal velocities of 0.8-1.2~km\,s$^{-1}$ just outside the sunspot with a monotonic decrease to a mean value of 180~m\,s$^{-1}$ at the outer edge of the moat cell where the noise level is reached. Mechanisms driving the moat flow are a controversial topic in solar physics. The non-existence of the moat flow when a sunspot lacks a penumbra and its transport function of magnetic features across the unmagnetised field to the surrounding network leads to the suggestion of the moat flow as an extension of the Evershed flow (\citealt{2008ApJ...679..900V}). While observations of a moat flow after the penumbra has dissolved (e.g., \citealt{2007ApJ...671.1013D,2012A&A...538A.109V,2013CEAB...37..435B}) and the fact that the Evershed flow is magnetised whereas the moat flow is, besides small magnetic features crossing it, an unmagnetised flow, contradicts this picture. In this contribution, to study the evolution of the moat flow, we analyse the flow field around sunspots during sunspots decay while the penumbra dissolves. Here we describe our analysis method, based on that by \citet{2013A&A...551A.105L}, yet, including the motion of the sunspots across the solar disc.
%%%%%%%%%%%%%%%%%%%%%%
\section{Data analysis}
Data provided by the satellite Solar Dynamics Observatory (SDO) allow for continuous observations of the solar disc and therefore to follow the evolution of an active region over several days. For the investigation of the moat flow, sub-frame data of the full disc with a field of view (FOV) of 500 x 500 pixel and a cadence of 720 s (12 minutes) provided by the webpage of the HMI/AIA Joint Science Operations Center (JSOC) are selected according to their NOAA. The sub-frame data are already tracked by the location of the active region. Intensity maps recorded by the Helioseismic and Magnetic Imager (HMI) are used to localise the sunspot in the FOV. Line-of-sight (LOS) magnetograms allow the observation of magnetic structures around the sunspots. The main analysis of the velocity profile and amplitude of the flow field is done by using Doppler maps recorded by HMI.
%%%%%%%%%%%
\subsection{Reduction of Doppler maps}
To be able to measure the flow velocities around sunspots, the Doppler maps have to be corrected from various effects. A first correction is done by using 720\,s-cadence data, which are provided by JSOC as low-noise data. In addition, the velocity pattern caused by the Sun's differential rotation, the centre-to-limb variation in the convective blueshift and a residual pattern caused by instrumental effects have to be subtracted from the selected Doppler maps. For each of these velocity patterns, a model is created following \citet{2013A&A...551A.105L}. The analysed active regions were observed in the years 2013 and 2014. As a new correction of the HMI Doppler maps was introduced after the Venus transit in 2012, new velocity maps of the mentioned velocity patterns had to be generated. We used 62 full-disc Doppler maps taken from 29 May to 17 June 2013, 30 November to 20 December 2013 and 29 May to 17 June 2014, by taking advantage of the inclination angle of $B_{0}\approx0^{\circ}$ for these time periods. The residual pattern is obtained after subtracting the centre-to-limb variation in the convective blueshift and the differential rotation from the averaged Doppler maps. Finally, the adjustment of the single sub-frame Doppler maps is done as described by \citet{2013A&A...551A.105L}. The Doppler maps are corrected from the individual radial and tangential velocity components of the satellite. The mentioned models of the velocity patterns allow for an individual reduction of the data by taking into account the inclination and the location of the FOV on the solar disc. To average out the effect of p-modes and reduce the noise level, 3\,h time averages (out of 15 Doppler maps) are created. This leads to a data set of 8 images per day for each target. This operation is in agreement with the findings of \citet{2013A&A...551A.105L}, who found a mostly constant moat flow in a time range of 3\,h. For comparison, the intensity maps and LOS-magnetograms are also averaged over 3 hours.
%%%%%%%%%%%
\subsection{Localisation of the region of interest}
The localisation of the centre of the sunspot in the 3 hour time averaged data is done by determining the centre of gravity in the intensity maps. A new FOV of 200\,x\,200\,pixel with the spot in the centre of the image is cropped out of the original ones. By using the intensity threshold of the spot compared to the surrounding region, the area of the spot is localised. The extension of the spot is defined through the radius, which is computed as the maximum distance between the outer contour line of the penumbra and the centre of the FOV. When the spot disappears in the intensity map, due to decay, a manual localisation of the vanished spot has to be done in order to investigate the flow in the appropriate region beyond the spot's existence. This is done by localising the position by eye, taking into account of the remaining magnetic flux in the LOS magnetograms. The selected coordinates are defined as the centre of the flow field. A FOV of 200\,x\,200\,pixel with the localised position in the centre of the image is then cropped out. The obtained coordinates are then used to localise the spot in the Doppler maps.
%%%%%%%%%%
\subsection{Analysis of the flow field}
\begin{figure}
\centering
\includegraphics[width=0.9\textwidth]{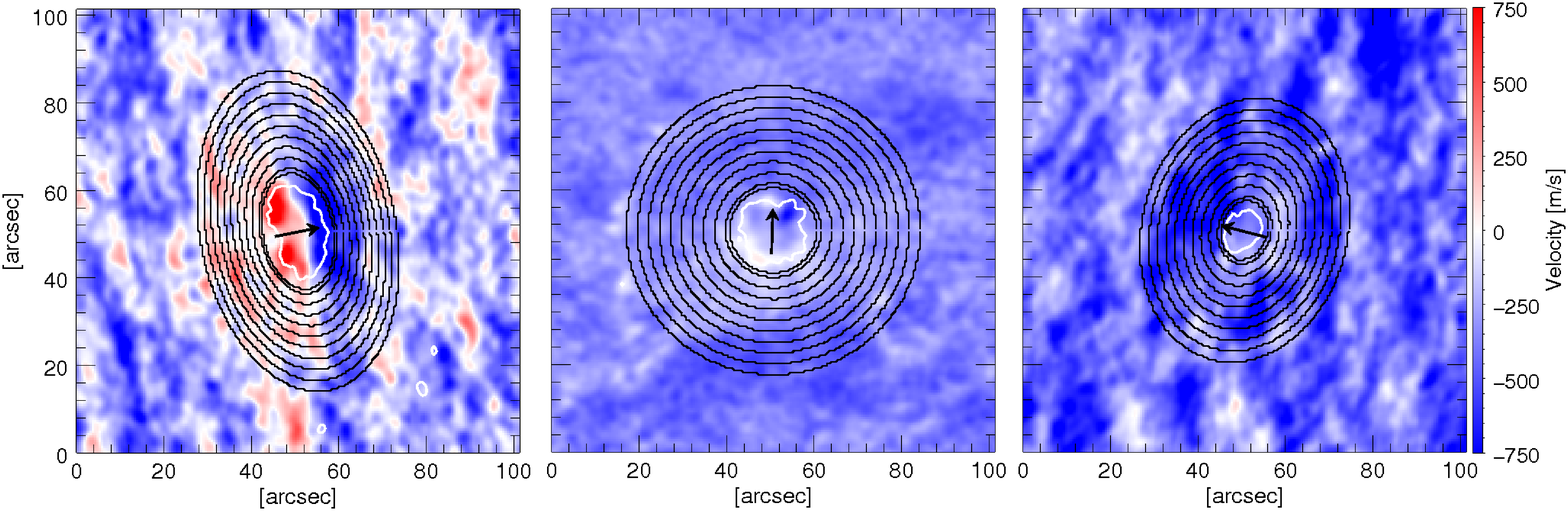}
\includegraphics[width=0.95\textwidth]{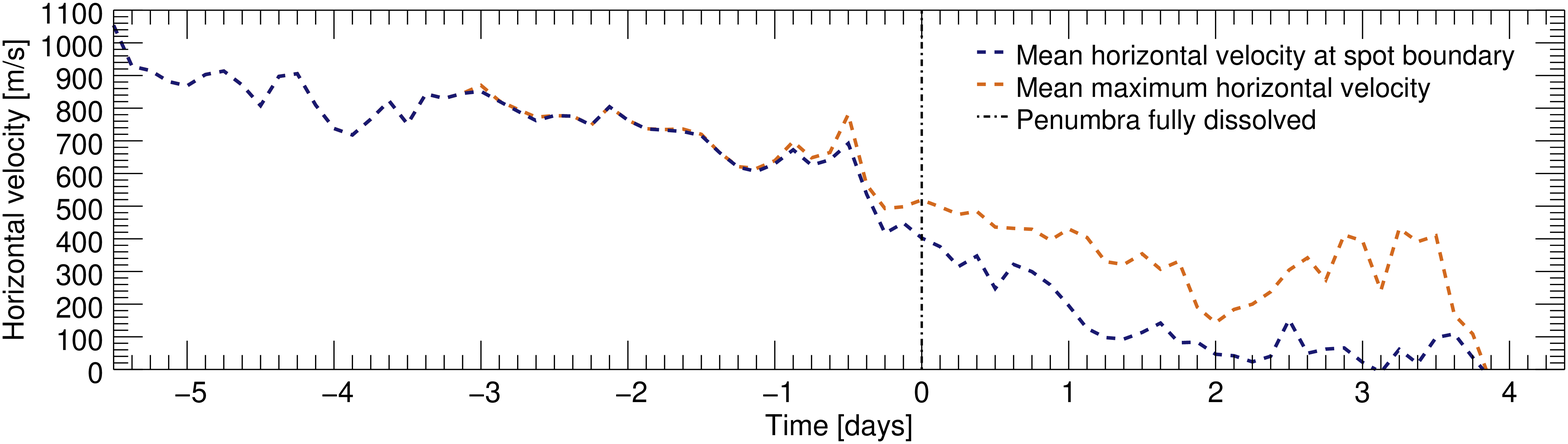}\caption{{\em Upper panels:} Examples of 3h-averaged Doppler maps of AR11841 for three different evolutionary stages and heliocentric angles: 39.86$^{\circ}$E (left), 10.83$^{\circ}$W (middle) and 39.63$^{\circ}$W (right). {\em Lower panels:} Mean LOS horizontal velocity around the 8 observed sunspots during their decay. The loss of the penumbra (vertical line) is set as a common point in time.}
\label{ex_fig}
\end{figure}

For the analysis of the flow field, the calibrated and 3\,hour averaged Doppler maps are used. Due to the spherical surface of the Sun, the (horizontal) outflow around the sunspot -- as well as the Evershed flow -- becomes visible as redshifted velocities away from disc centre and blueshifted velocities in the direction of disc centre, as it can be seen in Figure \ref{ex_fig}, upper-most left panel.
%%%
Following the analysis of \citet{2013A&A...551A.105L}, the applied method determines azimuthally averaged flow properties (\citealt{2000A&A...358.1122S}). The sunspots are assumed to be circular and have axially symmetrical flow fields. By applying this method, for each Doppler map, the LOS velocity values along ellipses with different radii $R\,=\,R_{spot}+r_{F}$ ($R_{spot}$: radius of the sunspot, $r_{F}$: distance of the ellipse to the spot boundary with $r_{F}\ge3\,pixel$ in one-pixel-steps) around a sunspot are extracted. The data array of each ellipse in the Doppler maps contains the LOS velocity, $\upsilon^{LOS}(R,\phi)$, as a function of the azimuth $\phi$. The velocity is read out, by forcing $\phi=90^{\circ}$ pointing towards solar disc centre. Similarly to \citet{2013A&A...551A.105L}, the LOS horizontal velocity component, $\upsilon^{LOS}_h(R)$, can be obtained as the amplitude of a sine fit of the LOS velocity. The LOS vertical flow component, $\upsilon^{LOS}_v(R)$, is obtained xas the offset of the fit, which depends strongly on the foregoing calibration of the data, which is influenced by several assumptions. Therefore, an accurate value of the vertical velocity component of the moat flow, for a meaningful analysis, cannot be obtained. The elliptical shape of the sunspot is due to the foreshortening effect at a certain heliocentric angle, $\theta$, off-disc centre and changes into a more circular shape as the spot crosses the $0^{\circ}$--meridian, as it can be seen in Figure \ref{ex_fig}, middle. We study decaying sunspots, therefore, they undergo significant changes in morphology. These asymmetries are taken into account in the analysis of the resulting flow profiles, e.g., by studying the magnetic structure of the sunspots and surroundings in LOS magnetograms. The measured LOS horizontal flow velocities on the solar surface also depend on the heliocentric angle, $\theta$. Thus \begin{equation}\upsilon_h(R)=\frac{\upsilon^{LOS}_h(R)}{\sin(\theta)}\label{vh}.\end{equation}This projection effect has to be taken into account in order to compare the obtained velocities of the flow around a sunspot for different positions during its passage over the solar disc.

Mean horizontal velocities measured during the evolution of the 8 observed active regions are displayed in Fig.\,\ref{ex_fig}, lower-most panel. They refer to measurements at the spot boundary ({\em blue}) and the maximum values found around sunspots ({\em orange}). The loss of penumbra is set as a common point in time (vertical line). Both profiles clearly detach at the time when penumbra dissolves. From this point in the sunspots' evolution on, a weaker (horizontal) velocity component around the vanishing spots still remain.
%%%%%%%%%%%%%%%%%%%%%%
\section{Discussion and conclusion}
The described analysis method is used to study the evolution of the flow field around 8 selected H-type sunspots, which decay as they cross the solar disc. 
We find that the flow field around stable sunspots with penumbra, the moat flow, is characterised by a horizontal velocity profile, which decreases with increasing distances to the spot boundary. We thus confirm the results by \citet{2013A&A...551A.105L}. Yet, most important, we also find that, as the penumbra dissolves and the sunspot decays, the velocity profile changes and lower horizontal velocity values than those for fully fledged sunspots are measured. This variation of the velocity profile of the flow field in the later stages of the sunspots' evolution suggests the action of a weaker flow taking over the moat flow, possibly a {\em supergranular flow}. We refer the reader to a forthcoming contribution on the comparison between the outflow measured around decaying sunspots and supergranular areas in the quiet-Sun. They shall shed some light onto a possible relation between the moat and supergranular flows.

% For non-BibTex:

\end{document}